\documentclass[final,5p,times,twocolumn]{elsarticle}

\usepackage{graphicx}
\usepackage{bm}
\usepackage{booktabs}
\usepackage{array}
\usepackage{hyperref}
\usepackage{cleveref}

\usepackage[pscoord]{eso-pic}

\newcommand{\placetextbox}[3]{
  \setbox0=\hbox{#3}
  \AddToShipoutPictureFG*{
    \put(\LenToUnit{#1\paperwidth},\LenToUnit{#2\paperheight}){\vtop{{\null}\makebox[0pt][c]{#3}}}%
  }%
}%

\usepackage{caption}
\usepackage{subcaption}

\usepackage{setspace}

\usepackage{lineno}

\begin{document}

\begin{frontmatter}

\title{Radiation Damage of $2 \times 2 \times 1 \ \mathrm{cm}^3$ Pixelated CdZnTe Due to High-Energy Protons}

\author[3]{Daniel Shy}
\author[2]{David Goodman}
\author[2]{Ryan Parsons}
\author[2]{Michael Streicher}
\author[2]{Willy Kaye}
\author[3]{Lee Mitchell}
\author[2]{Zhong He}
\author[3]{Bernard Phlips}

\address[3]{U.S. Naval Research Laboratory, 4555 Overlook Ave SW, Washington, DC 20375}
\address[2]{H3D, Inc., 812 Avis Dr, Ann Arbor, MI 48108, USA}


\begin{abstract}
Pixelated CdZnTe detectors are a promising imaging-spectrometer for gamma-ray astrophysics due to their combination of relatively high energy resolution with room temperature operation negating the need for cryogenic cooling. This reduces the size, weight, and power requirements for telescope-based radiation detectors. Nevertheless, operating CdZnTe in orbit will expose it to the harsh radiation environment of space. This work, therefore, studies the effects of $61 \ \mathrm{MeV}$ protons on $2 \times 2 \times 1 \ \mathrm{cm}^3$ pixelated CdZnTe and quantifies proton-induced radiation damage of fluences up to  $2.6 \times 10^8 \ \mathrm{p/cm^2}$. In addition, we studied the effects of irradiation on two separate instruments: one was biased and operational during irradiation while the other remained unbiased. Following final irradiation, the $662 \ \mathrm{keV}$ centroid and nominal $1\%$ resolution of the detectors were degraded to $642.7 \ \mathrm{keV}, 4.9 \% \ ( \mathrm{FWHM})$ and $653.8 \ \mathrm{keV}, 1.75 \% \ (\mathrm{FWHM})$ for the biased and unbiased systems respectively. We therefore observe a possible bias dependency on proton-induced radiation damage in CdZnTe. This work also reports on the resulting activation and recovery of the instrument following room temperature and $60^{\circ}\mathrm{C}$ annealing.

\end{abstract}


\end{frontmatter}


\section{Introduction}
\label{sec1}

\placetextbox{0.5}{0.05}{\large\textsf{Distribution Statement A: Approved for public release. Distribution is unlimited.}}%


Solid state detectors present an advantage over scintillation-based detectors for use in radiation detection applications due to their generally superior energy resolution~\cite{knoll}. This advantage stems from its direct-conversion process as semiconductors generally require less energy to produce an information carrier and result in a smaller fano factor. This results in better energy resolution as they are less limited by statistical fluctuation when compared to their scintillator counterparts. CdZnTe (CZT), a wide band gap, room temperature semiconductor, is a promising high-resolution gamma/x-ray spectrometer for space deployments.

CZT has a significant heritage of being used in space experiments. It has flown in several balloon flights~\cite{cztBalloon,cztBalloon2, italianCZT} as well as flying in space on instruments such as the BAT-Swift Observatory~\cite{swift}, AstroSat~\cite{czti}, Dawn~\cite{DAWN_CZT}, and ASIM~\cite{asim}. It is also considered for future missions such as AMEGO~\cite{amego}, GECCO~\cite{gecco}, and ASTENA~\cite{ASTENA}.

If the instrument is placed in low earth orbit, it will likely pass through the Southern Atlantic Anomaly (SAA) to some varying degree that depends on orbit inclination~\cite{SAA}. There, it will experience a high flux of trapped energetic particles. The space radiation environment also contains galactic cosmic rays (mostly energetic protons and alphas at $10 \ \mathrm{GeV/nucleon}$) and solar-based radiation from solar activity~\cite{spaceRadiation}. This radiation environment can severely degrade the performance of gamma-ray spectrometers~\cite{LEBRUN2005323}.

In this work, we study the damage induced by $61 \ \mathrm{MeV}$ protons on CZT. Although the proton flux in the SAA is represented by a power law in between $10-100 \ \mathrm{MeV}$~\cite{protonSAAMap}, $61 \ \mathrm{MeV}$ protons are still within that range, and therefore somewhat illustrative of the actual environment. Sec.~\ref{sec:prevWork} discusses previous work concerning radiation damage in CZT. Sec.~\ref{sec:methods} is the methods section describing the CZT hardware, the irradiation facility, and the irradiation plan. Sec.~\ref{sec:damageResults} presents the results from the proton irradiations and Sec.~\ref{sec:recovery} presents the detector's recovery with annealing. Sec.~\ref{sec:activiation} offers an analysis of the observed activation.

\section{Previous Work on High-Energy Proton Irradiation With CZT}
\label{sec:prevWork}

Several radiation damage studies on CZT were conducted with protons on the order of several hundred $\mathrm{MeV}$. Varnell et al. irradiated $2-3 \ \mathrm{mm}$ planar CZT with $200 \ \mathrm{MeV}$ protons~\cite{radDamage199}. They observed that as the dose increased, the spectral gain decreased proportionally due to the production of electron traps which decreases the mobility lifetime product $(\mu \tau _e)$ of electrons. With $3 \ \mathrm{mm}$ CZT, they observed a shift in the peak centroid with a fluence as low as $1\times 10^8 \ \mathrm{p/cm^2}$~\cite{603768}. That same group irradiated pixelated CdZnTe and observed a factor of 4.4 degradation in the $\mu \tau _e$ of the worst detector that experienced $5 \times 10^9 \ \mathrm{p/cm^2}$. Franks et al. provide an overview of damage in various room temperature semiconductors~\cite {FRANKS1995}. In CZT, they report that planar $3 \ \mathrm{mm}$ thick CZT exposed to $200 \ \mathrm{MeV}$ experiences a two-fold loss in energy resolution after a fluence of $5 \times 10^9 \ \mathrm{p/cm^2}$ but little effect in $2 \ \mathrm{mm}$ thick devices. Moreover, when testing $2 \ \mathrm{mm}$ planar devices they observed larger resolution losses in an unbiased detector when compared to the biased case. The electric field strength, or voltage bias, is not reported in the cited work.

A few years later, Kuvvetli et al. irradiated a $2.7 \ \mathrm{mm}$ thick drift strip detector with $30 \ \mathrm{MeV}$ protons~\cite{KUVVETLI_CZT_Damage}. The first irradiation of $2.8 \times 10^8 \ \mathrm{p/cm^2}$ resulted in a noticeable decline in $\mu \tau _e$. They were able to recover the detector after annealing at $100^{\circ} \mathrm{C}$ for a total time of 22 hours.

Zanarini et al. irradiated $1 \ \mathrm{mm}$ thick CZT with $2\ \mathrm{MeV}$ protons and did not observe any gain shift or resolution change even with an exposure of $10^{11} \ \mathrm{p/cm^2}$~\cite{2MeVCZT}.

It is worth noting that the aforementioned detectors were relatively thin, on the order of millimeters. Bolotnikov et al. irradiated larger volume $8 \times 8 \times 32 \ \mathrm{mm}^3$ virtual Frisch-grid detectors~\cite{BNLProton} with $150 \ \mathrm{MeV}$ protons. Irradiated from the `side', they did not notice any polarization during acquisition with low proton flux of $160 \ \mathrm{p/cm^2/s}$. During the high flux damage study, they did not notice any gain shift with fluences up to $10^6 \ \mathrm{p/cm^2}$. However, with a fluence of $4 \times 10^7 \ \mathrm{p/cm^2}$, they began noticing a small gain shift of less than $2\%$. After $10^{10} \ \mathrm{p/cm^2}$, the detectors stopped responding. Nevertheless, the detectors are fully recovered after annealing at $65 \ ^{\circ}\mathrm{C}$ and then $80 \ ^{\circ}\mathrm{C}$ for 3 and 2 weeks respectively. During irradiation, their detectors were biased with a field of $115.6 \ \mathrm{V/mm}$.

\section{Materials and Methods}
\label{sec:methods}

\subsection{CdZnTe Hardware}

This work uses two M400 CZT modules~\cite{m400, m400Paper} manufactured by H3D, Inc. The module utilizes four crystals, each with a volume of $2\times 2 \times 1 \ \mathrm{cm}^3$. The anode contains an array of $11 \times 11$ pixels with a pitch of $1.72 \ \mathrm{mm}$ with a planar cathode. This allows for the 3D reconstruction of interaction positions. The crystals are arranged in a $2 \times 2$ array with the cathodes facing towards the incident proton beam.

In this study, the manufacturer calibrated the system beforehand at a different location. Unless stated otherwise, we present the results using the aforementioned calibration, which allows for the observation of the gain shift and resolution degradation during irradiation. To calibrate the CZT system a flood irradiation is taken~\cite{andythesis}. The cathode-to-anode ratio is taken to estimate the depth of interaction. Next, for a given pixel, the events are separated into multiple depth bins.  The raw, uncalibrated photopeak amplitude is measured on a voxel-by-voxel basis and used to correct voxelized spectra collected in subsequent measurements.

\subsection{Proton Beam Characteristics}

We used the cyclotron at the Crocker Nuclear Laboratory on the University of California Davis’ campus~\cite{DavisBeam}. The cyclotron can produce monoenergetic protons at various energies. This experiment utilizes a $67.5 \ \mathrm{MeV}$ beam that passes through a Kapton window, tantalum, and aluminum diffuser. The circular beam measures $6 \ \mathrm{cm}$ in diameter, enough to uniformly irradiate the entire M400 module. The uniformity of the beam is reported in~\cite{DavisBeam} and varies less than $4\%$ in the inner $2 \ \mathrm{cm}$ radius.

\subsubsection{Simulation of Proton Response in CdZnTe}

\begin{figure}[h!]
  \centering
  \includegraphics[trim={0cm 0cm 0cm 0cm}, clip, width=\linewidth]{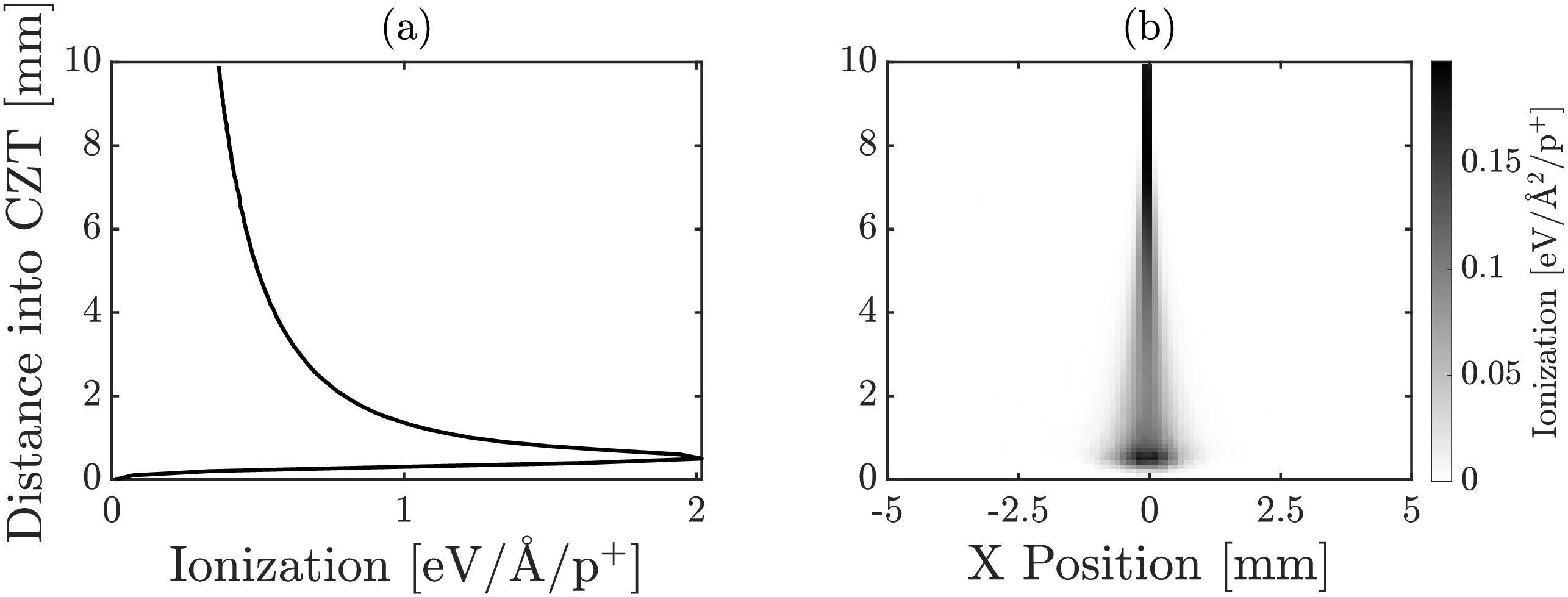}
  \caption{Simulated ionization along the path of the proton. (a) Shows the ionization projected onto the depth of the detector while (b) shows a 2D distribution of ionization. Zero depth in this plot reflects the anode.}
  \label{fig:srim}
\end{figure}

Using a SRIM simulation~\cite{SRIM} that accounts for the attenuation of intervening air and M400 housing we estimate a mean proton energy of $61.0 \ \mathrm{MeV}$ impinging on the crystal. Fig.~\ref{fig:srim}a shows the average energy loss of incident protons, or the electronic energy loss~\cite{SRIM}, as a function of the depth in the CdZnTe crystal using SRIM. The estimated proton range is $9.5 \ \mathrm{mm}$ with a straggling of $227 \ \mu \mathrm{m}$. Fig.~\ref{fig:srim}b shows the two-dimensional ionization distribution of the simulated proton pencil beam.

\subsection{Experimental Setup and Irradiation Plan}

Two M400 modules were irradiated; one was biased and unbiased respectively during irradiation. The biased module operated at $2000 \ \mathrm{V}$ and was acquiring spectra the entire time. The unbiased module was powered-off which disabled everything including the front-end electronics and embedded computer.

\begin{figure}[h!]
  \centering
  \includegraphics[trim={0cm 0cm 0cm 0cm}, clip, width=\linewidth]{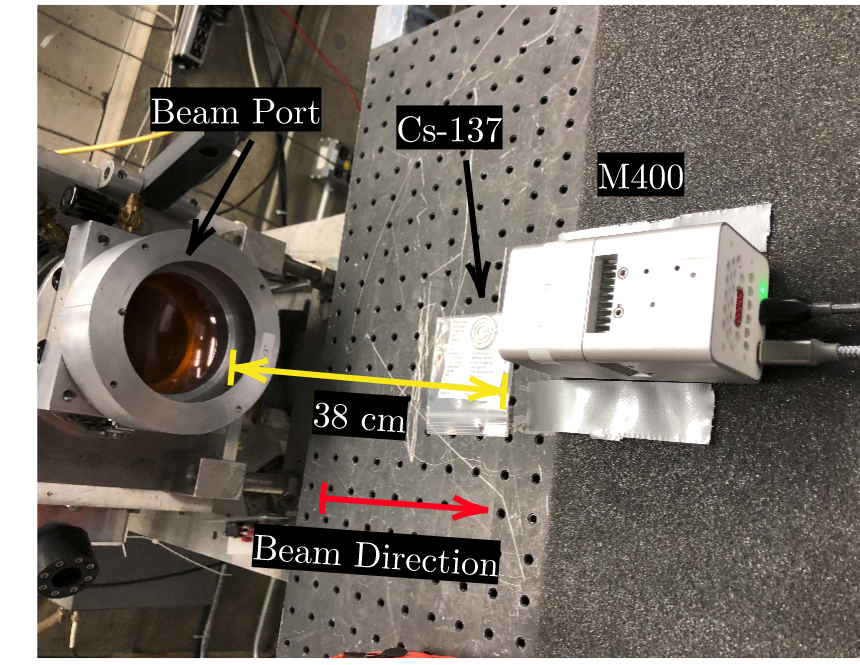}
  \caption{Experimental setup at the irradiation facility. Detectors were place on top foam to distance them from high-Z objects.}
  \label{fig:expSetup}
\end{figure}

Each detector was iteratively irradiated to the prescribed fluence followed by measurement of a Cs137 check source. The source was kept in the target room throughout the experiment and placed on the cathode side - slightly off axis. The unbiased detector was turned on following irradiations to collect 137Cs spectra and then powered off. Fig.~\ref{fig:expSetup} shows the experimental setup with the detector. 137Cs spectra were collected with accumulated fluences of up to $2.6 \times 10^8 \ \mathrm{p/cm^2}$. Using thin target approximation, this proton fluence is equivalent to  ${\sim}45$ rad(Si), or ${\sim}3$ years on the International Space Station for an external pallet on the Columbus module: that location expects $149 \ \mathrm{ mGy/yr}$ ($14.9 \ \mathrm{rad/yr}$)~\cite{ISSRad}. Note that the thin target approximation is not applicable in this situation, but is given a means of comparison.

\begin{figure}[h!]
  \centering
  \includegraphics[trim={0cm 0cm 0cm 0cm}, clip, width=0.65\linewidth]{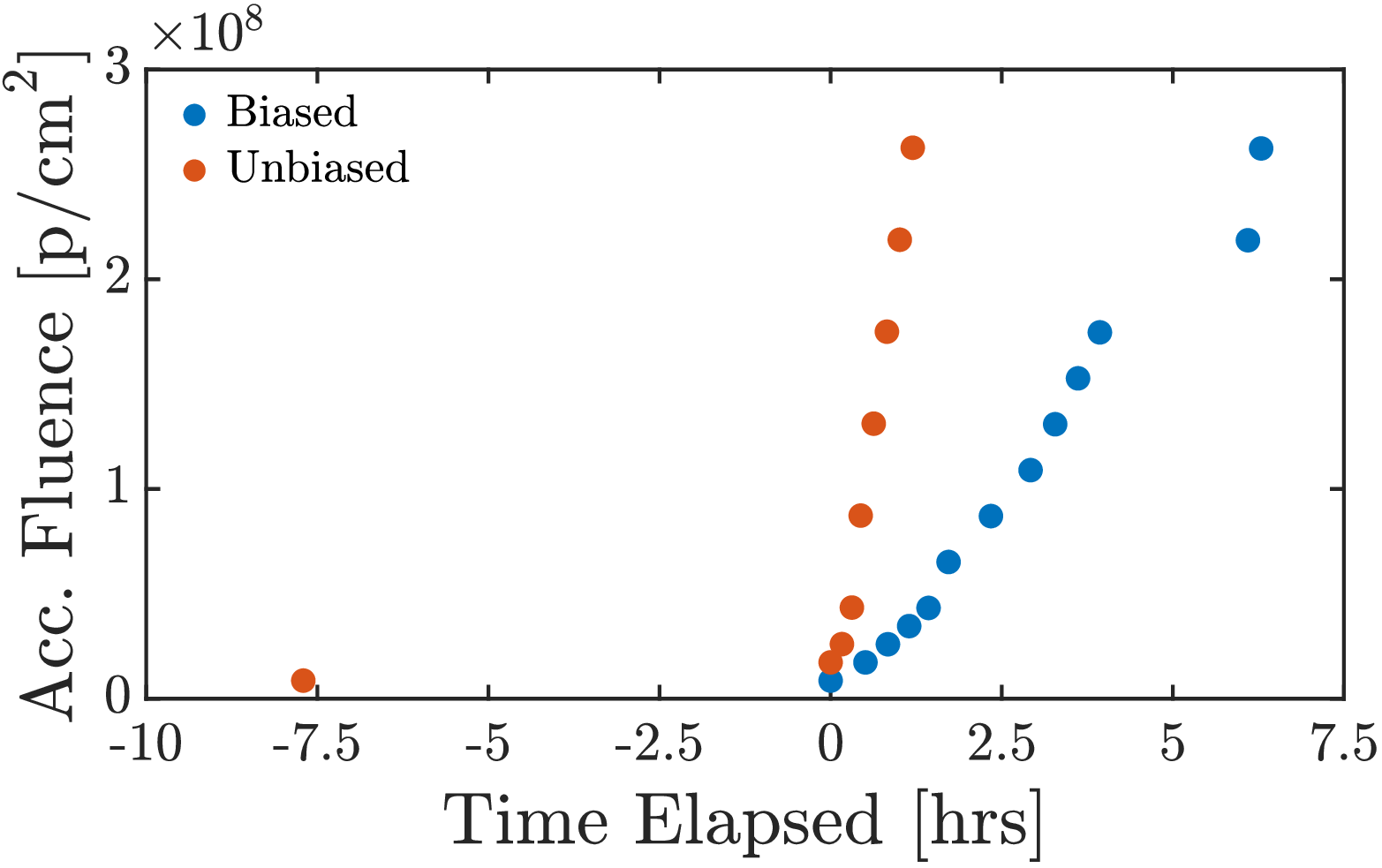}
  \caption{Irradiation history for the biased and unbiased detectors. The time elapsed signifies the time passed since the first irradiation for the biased detector. The unbiased detector is elapsed time since the second irradiation. This is due to the first irradiation in the unbiased detector taking place the night before the main experiment.}
  \label{fig:expIrradiationPlan}
\end{figure}

Fig.~\ref{fig:expIrradiationPlan} presents the accumulated proton dose as a function of time. The unbiased detector received its first irradiation the night before the rest of the experiment. Its date is therefore shifted such that the time elapsed is zeroed to time from the 2nd irradiation. That detector then received dosages in near-constant intervals. The biased detector also received consistent dosages except for the last two runs, which had a 2-hour gap.

\section{Proton Irradiation Results}
\label{sec:damageResults}

\begin{figure}[h!]
  \centering
  \includegraphics[width=\linewidth]{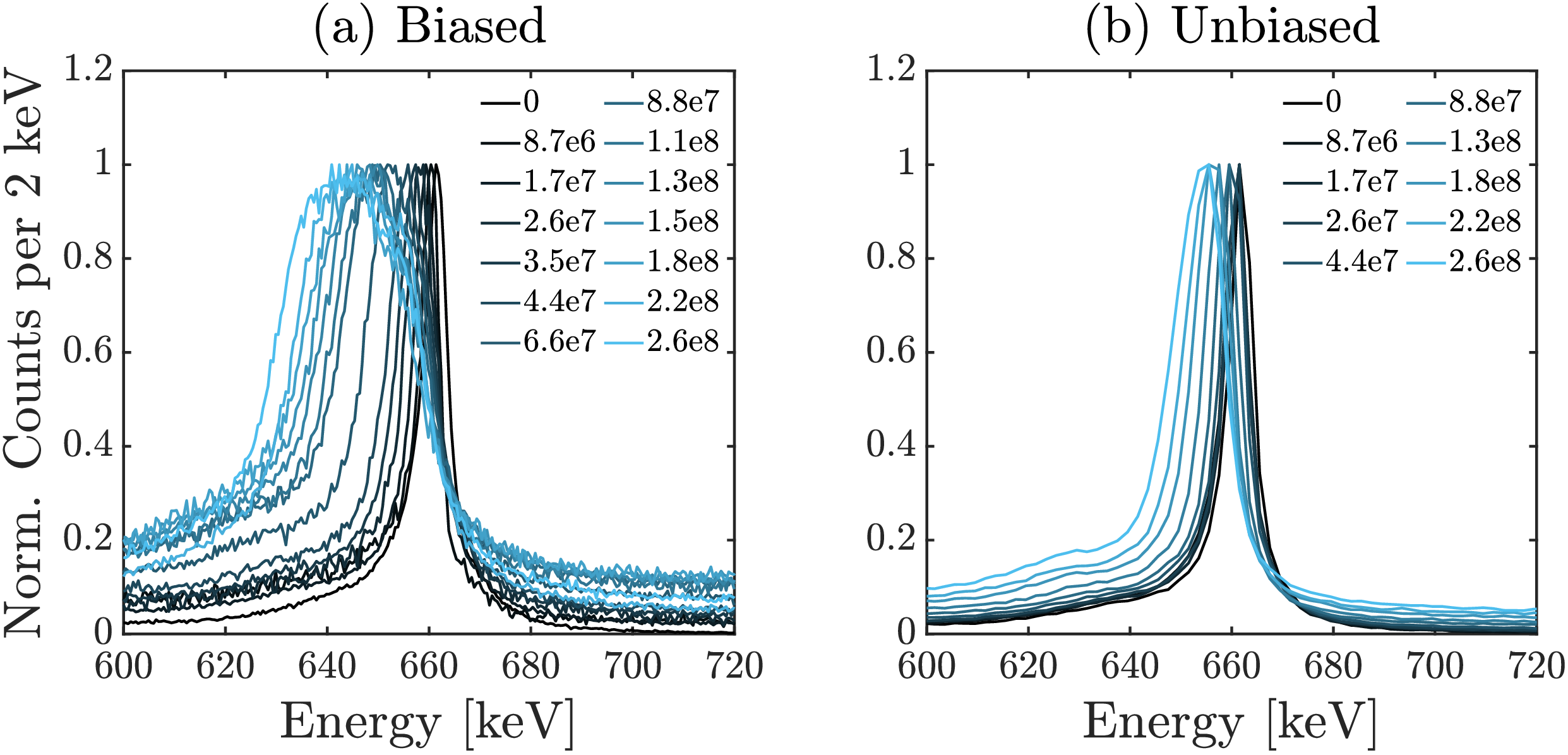}
  \caption{137Cs Spectra and its degradation at different accumulated fluences. (a) Presents the spectra for the biased detector while (b) presents the unbiased response.}
  \label{fig:spectraVsRad}
\end{figure}

\begin{figure}[h!]
  \centering
  \includegraphics[trim={0cm 0cm 0cm 0cm}, clip, width=\linewidth]{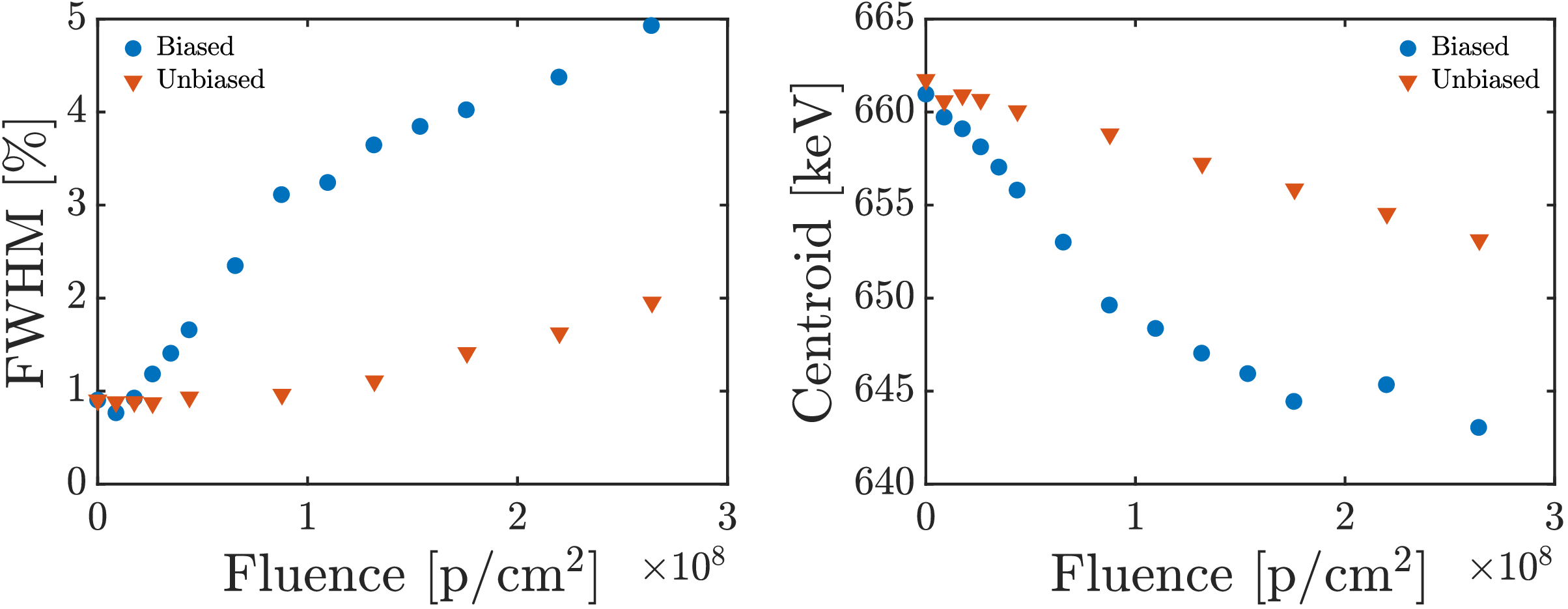}
  \caption{Quantitative features of the 137Cs peak measurements following different fluences for the biased and unbiased detector. (a) Presents the FWHM while (b) presents the peak centroid.}
  \label{fig:radiationDamage}
\end{figure}

Fig.~\ref{fig:spectraVsRad} plots the measured 137Cs spectra after a given accumulated irradiation for (a) biased and (b) unbiased detectors. The spectra present all the events recorded by the four crystals with no event cuts applied. This includes multi-interaction events. At first glance, the peak centroid decreases, and width increases with accumulated fluence. Moreover, qualitatively, the responses appear significantly different between the biased and unbiased cases. Next, in the most degraded spectrum, we observe a `flat topping' of the peak, which indicates a non-uniform response across depth. Fig.~\ref{fig:radiationDamage}a plots the full width at half maximum (FWHM) as a function of the accumulated fluence. After the full prescribed irradiation of $2.6 \times 10^8 \ \mathrm{p/cm^2}$, we observe over a factor of two poorer resolution in the biased case versus the unbiased case. The 0 accumulated fluence point is slightly off trend as it was measured at a different time and place than the rest of the data.

We observe similar behavior in Fig.~\ref{fig:radiationDamage}b, which plots the peak centroid as a function of the accumulated fluence. There, after $1.7 \times 10^8 \ \mathrm{p/cm^2}$, the centroid drifted $2.6\%$ in the biased detector compared to $0.8\%$ observed in the unbiased case. Note that the last two data points in the biased detector are slightly off-trend due to the 2-hour break between those sets (see Fig.~\ref{fig:expIrradiationPlan}).

\begin{figure}[h!]
  \centering
  \includegraphics[trim={0cm 0cm 0cm 0cm}, clip, width=\linewidth]{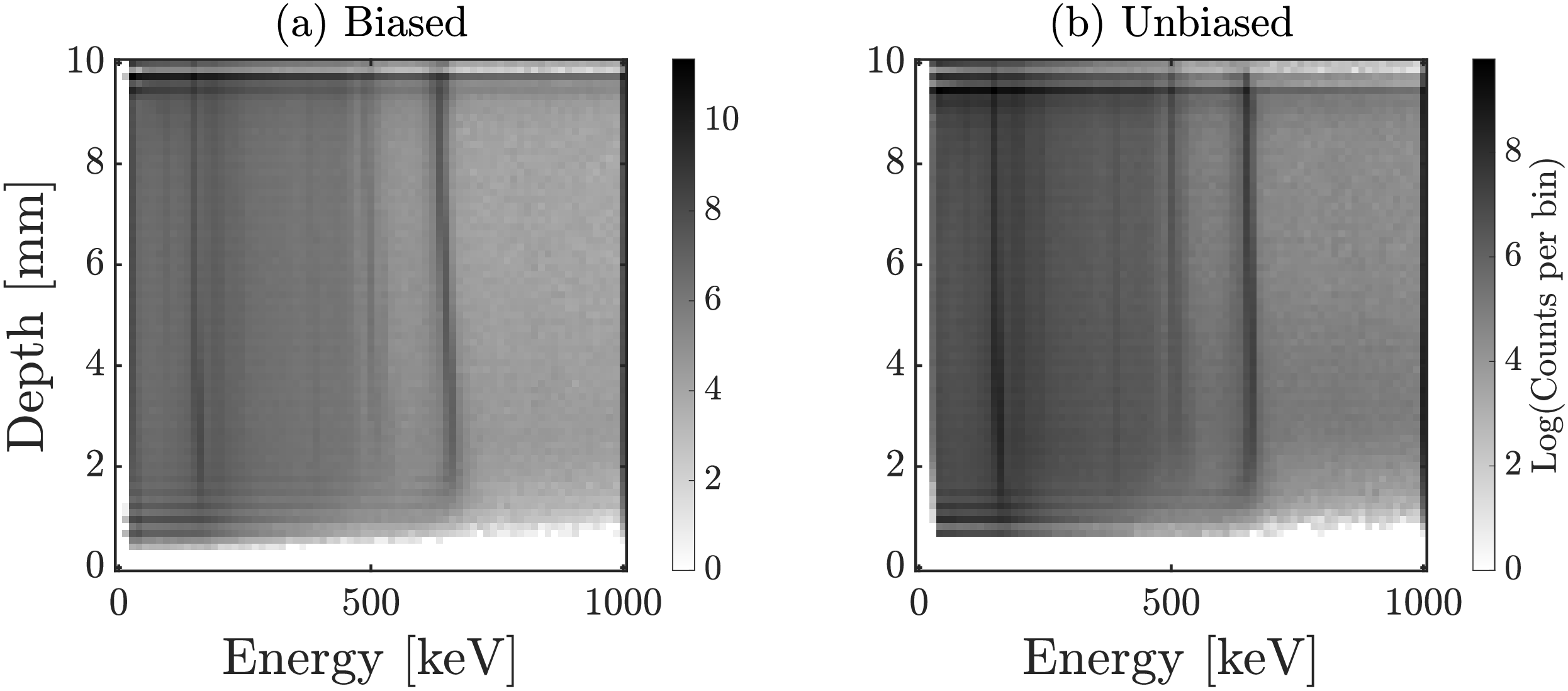}
  \caption{Depth-energy responses of a 137Cs measurement following $2.6 \times 10^8 \ \mathrm{p/cm^2}$ accumulated fluence for the (a) biased and (b) unbiased detectors. Zero depth represents the anode. Note the curvature of the 137Cs peak towards the anode as cathode side events must drift through more of the defective material and are susceptible to trappings.}
  \label{fig:bivariate}
\end{figure}

The bivariate plots between energy and depth from the cathode best show the difference between the two detector’s damage responses. Fig.~\ref{fig:bivariate} plots the summed 137Cs energy-depth response for all four crystals in a given bias state. The plots treat multi-interaction events as singles. In the plots, $0 \ \mathrm{mm}$ and $10 \ \mathrm{mm}$ represent the cathode and anode side respectively. The most intense vertical line is the gain-shifted $662 \ \mathrm{keV}$ line, while other lines represent internal activation. Following the $662 \ \mathrm{keV}$ peak along the depth, we see anode side interactions have a higher gain than cathode side events. This is due to the proton-induced traps introduced along the depth of the crystal. Moreover, the response is roughly flat along the cathode side until ${\sim} 5 \ \mathrm{mm}$ away from the anode. This follows the $d\mathrm{E}/dx$ the proton experiences as shown in Fig.~\ref{fig:srim}a where the ionization is roughly flat in the first ${\sim} 5 \ \mathrm{mm}$ from the cathode. Next, we observe a significant difference in the slope of the photopeak centroid with depth between the (a) biased and (b) unbiased cases, especially on the anode side.

\section{Recovery and Annealing Studies}
\label{sec:recovery}

This section presents the recovery of the CZT detectors from annealing at room temperature and baking in a $60 \ ^{\circ}\mathrm{C}$ oven. Unless stated otherwise, the displayed data uses a calibration taken pre-irradiation to display the recovery of the system relative to its original state.

\subsection{Annealing}

\begin{figure}[h!]
  \centering
  \includegraphics[trim={0cm 0cm 0cm 0cm}, clip, width=1\linewidth]{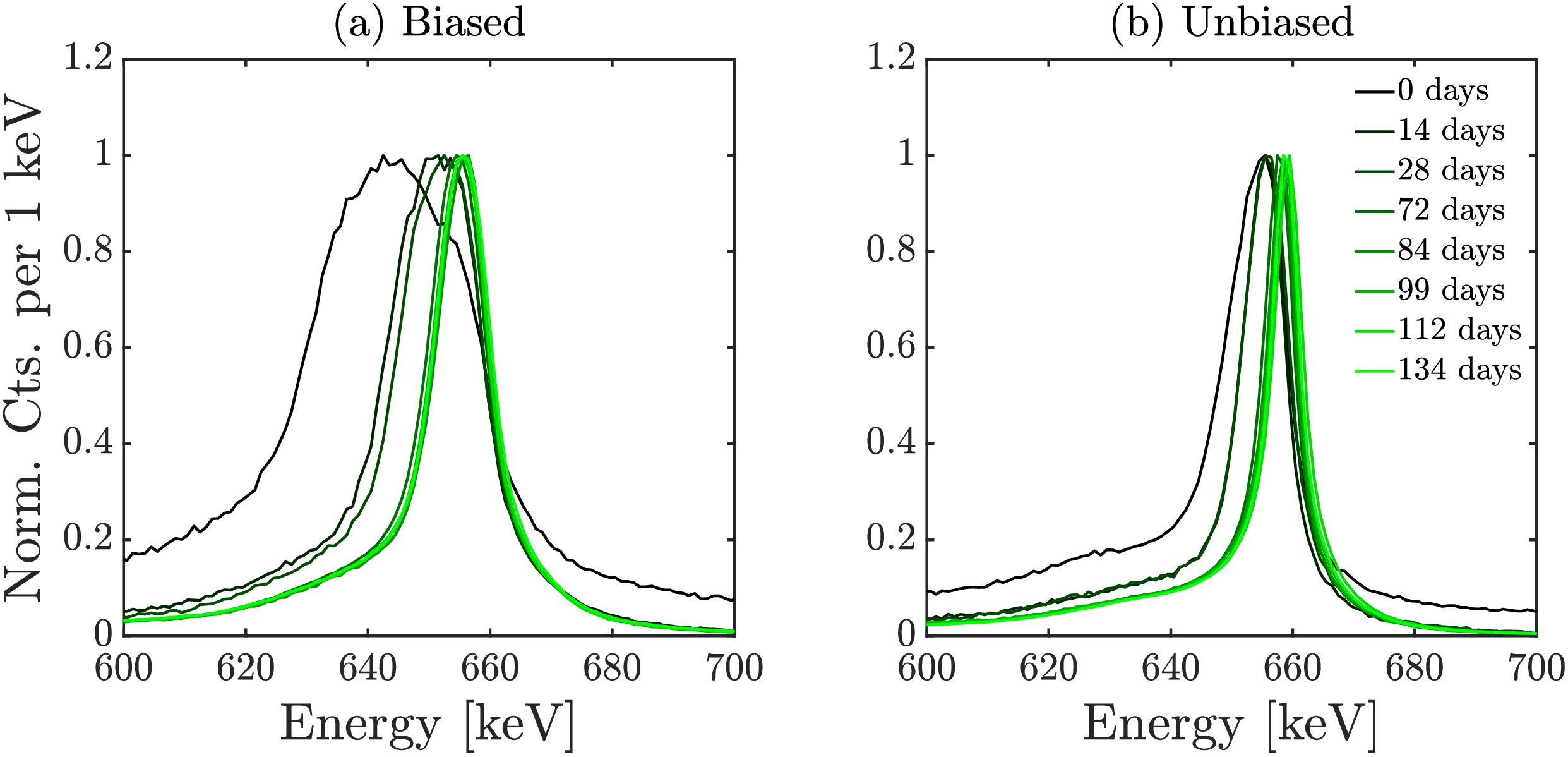}
  \caption{137Cs spectra showing recovery over time for the (a) biased and (b) unbiased case.}
  \label{fig:recoverySpectra}
\end{figure}

\begin{figure}[h!]
  \centering
  \includegraphics[trim={0cm 0cm 0cm 0cm}, clip, width=1\linewidth]{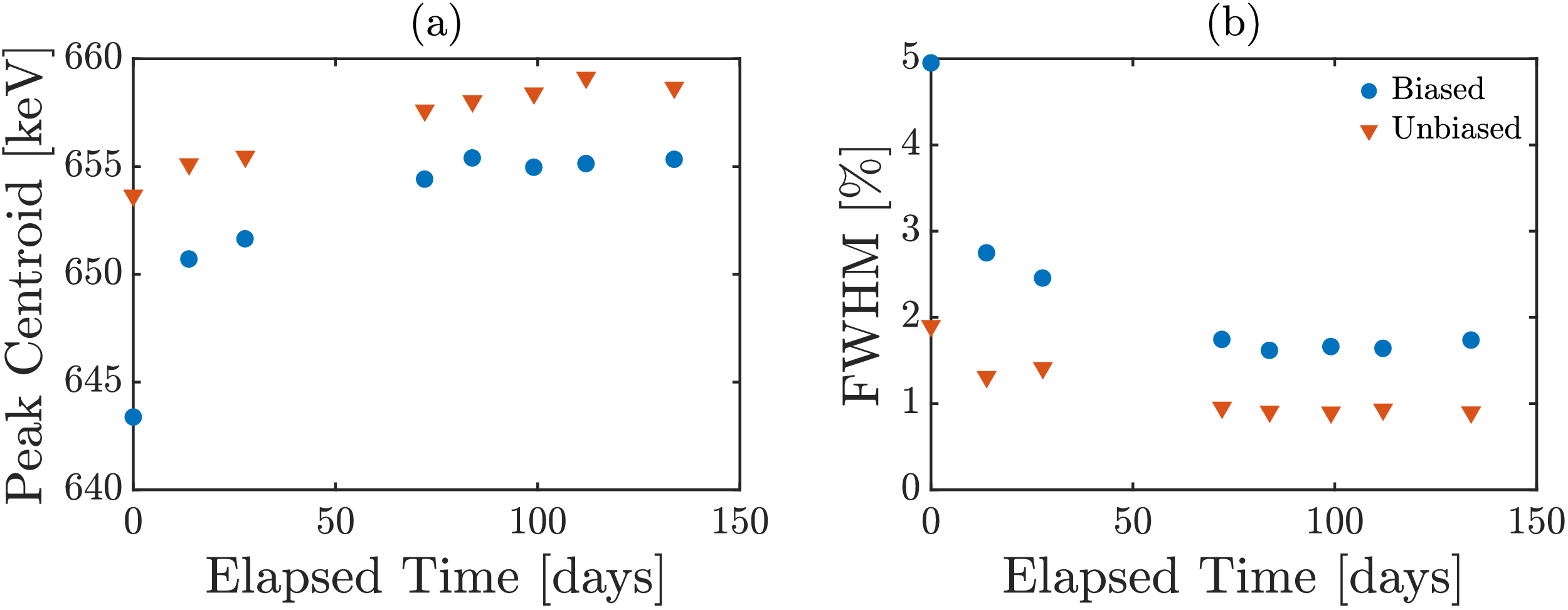}
  \caption{Recovery characteristics for the detectors annealing in room temperature. (a) plots the peak centroid over time while (b) plots the FWHM over time.}
  \label{fig:recoveryChar}
\end{figure}

Following irradiation, the CZT detectors were stored in an office environment. In roughly a two-week cadence, we took a 137Cs measurement to study the recovery of the detectors. Fig.~\ref{fig:recoverySpectra} plots the spectra for the (a) biased and (b) unbiased detectors. Fig.~\ref{fig:recoveryChar} plots the (a) peak centroid as a function of time while (b) plots the FWHM recovery. The plot shows that the first two weeks of annealing at room temperature resulted in the most significant change and reached a plateau at around 75 days. The resolution of the unbiased system plateaued to its pre-irradiated state of $\sim 1 \%$, while the biased system held steady at $\sim 1.7 \%$. Room temperature annealing in CZT has been previously observed in~\cite{CZTneutronDamage}.

\begin{figure}[h!]
  \centering
  \includegraphics[trim={0cm 0cm 0cm 0cm}, clip, width=1\linewidth]{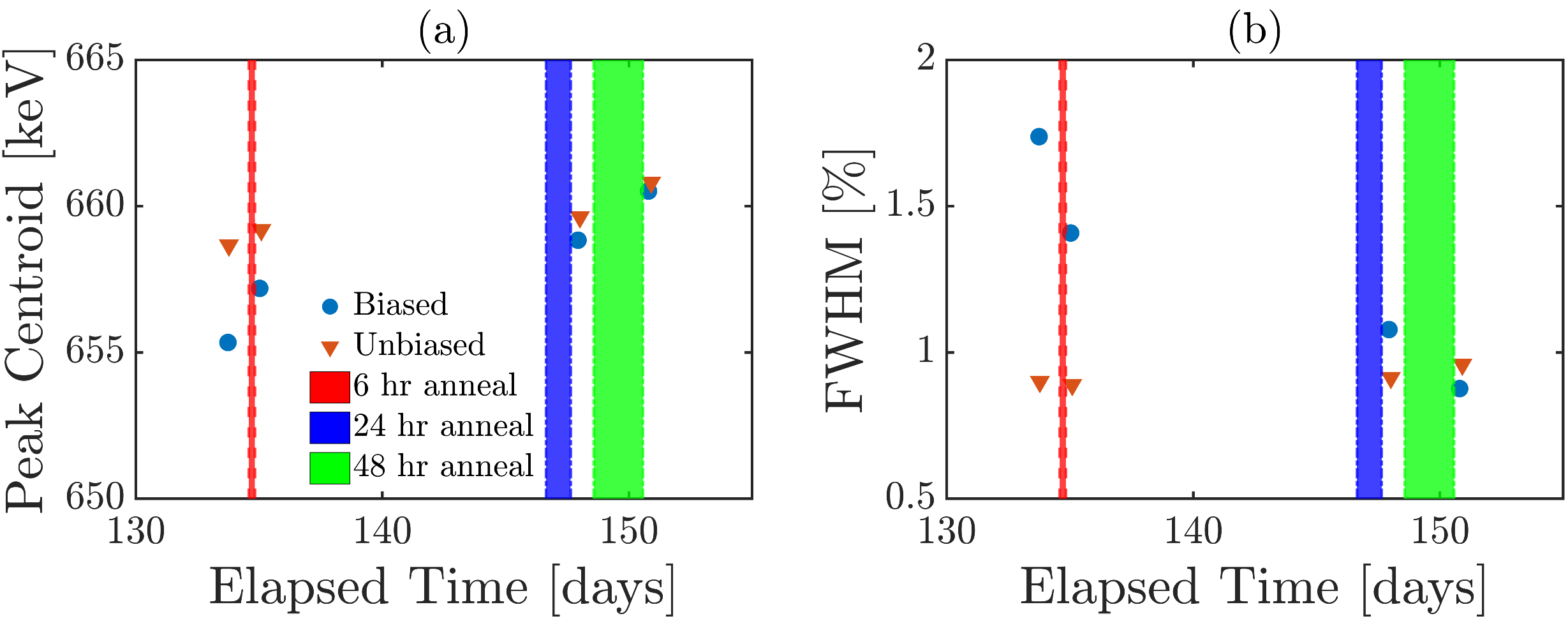}
  \caption{Recovery of the CZT detectors over several annealing sessions. The shaded time frames indicate periods in which the detectors were in the oven. (a) Plots the recovery of the peak centroid while (b) plots the FWHM.}
  \label{fig:annealRecovery}
\end{figure}

After 135 days both detectors were annealed in a $60 \ ^{\circ}\mathrm{C}$ three times. The sessions lasted 6, 24, and 48 hours. We chose the temperature due to several temperature limits not associated with the crystals. The detectors were then allowed to return to room temperature after which a 137Cs measurement was taken. Figure~\ref{fig:annealRecovery} plots the performance of the system after each annealing step. It shows that 78 hours in the oven was sufficient to recover the biased system to its pre-irradiated resolution. The unbiased system did not show any significant change as its resolution already recovered. Table~\ref{tab:anneal} summarizes this recovery.

\begin{table}[]

\centering
\caption{Summary of the radiation damaged CZT for the biased and unbiased detector after room temperature (RT) annealing and $60 \ ^{\circ}\mathrm{C}$ annealing. The data in the field is presented as [centroid (keV), FWHM resolution ($\%$)].}
\label{tab:anneal}
\begin{tabular}{l c c}
\toprule

 Condition           & Unbiased           & Biased             \\
             \hline
             \hline
0 day RT anneal      & {[}653.7, 1.9\%{]} & {[}643.3, 4.9\%{]} \\ 
134 days RT anneal    & {[}658.7, 0.9\%{]} & {[}655.3, 1.7\%{]} \\ 
6 hr $60\ ^{\circ}\mathrm{C}$ anneal  & {[}659.2, 0.9\%{]} & {[}657.2, 1.4\%{]} \\ 
24 hr $60\ ^{\circ}\mathrm{C}$ anneal & {[}659.6, 0.9\%{]} & {[}658.8, 1.1\%{]} \\ 
48 hr $60\ ^{\circ}\mathrm{C}$ anneal    & {[}660.8, 1.0\%{]}   & {[}660.5, 0.9\%{]} \\
\bottomrule
\end{tabular}
\end{table}

\subsection{Application of Calibration}

\begin{figure}
     \centering
     \begin{subfigure}[b]{0.4\textwidth}
          \centering
         \includegraphics[trim={0.5cm 0.1cm 0.2cm 0cm}, clip,height=0.75\textwidth]{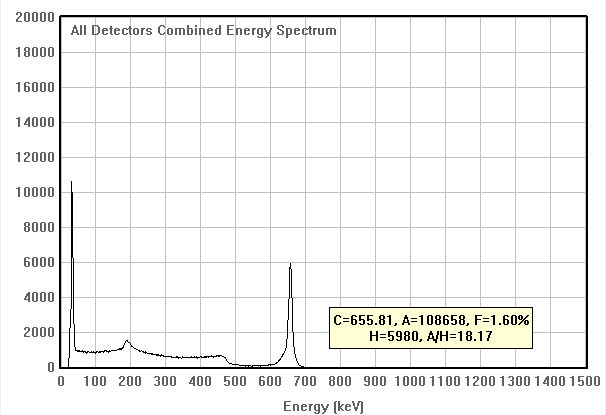}
         \caption{Pre-calibration}
         \label{fig:celinePre}
     \end{subfigure}
     \hspace{1em}
     \begin{subfigure}[b]{0.4\textwidth}
         \centering
         \includegraphics[trim={0.5cm 0.1cm 0.2cm 0cm}, clip,height=0.75\textwidth]{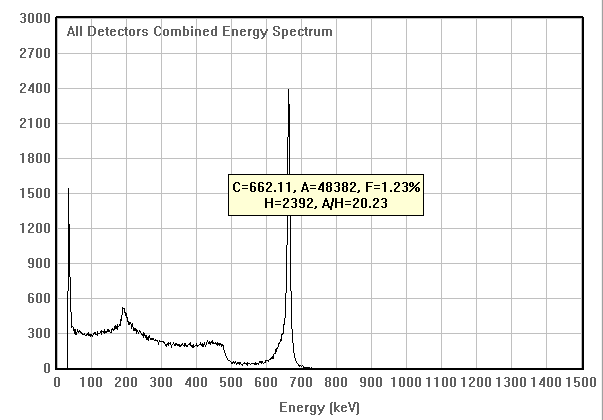}
         \caption{Post-calibration}
         \label{fig:celinePost}
     \end{subfigure}

        \caption{The biased detector Cs-137 spectrum (a) pre- and (b) post-calibration.}
        \label{fig:celineCalibration}
\end{figure}

\begin{figure}
     \centering

     \begin{subfigure}[b]{0.4\textwidth}
          \centering
         \includegraphics[trim={0.5cm 0.1cm 0.2cm 0cm}, clip,height=0.75\textwidth]{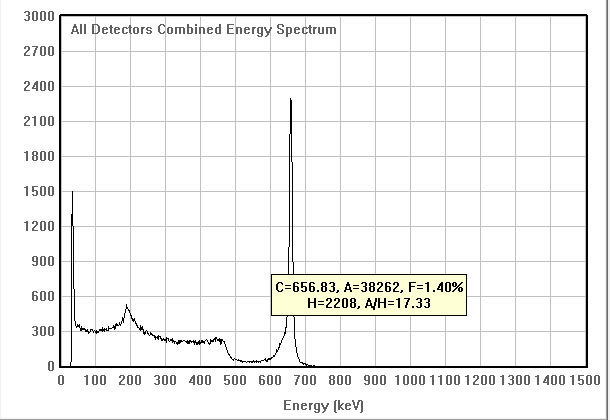}
         \caption{Pre-calibration}
         \label{fig:monizPre}
     \end{subfigure}
     \hspace{1em}
     \begin{subfigure}[b]{0.4\textwidth}
         \centering
         \includegraphics[trim={0.5cm 0.1cm 0.2cm 0cm}, clip,height=0.75\textwidth]{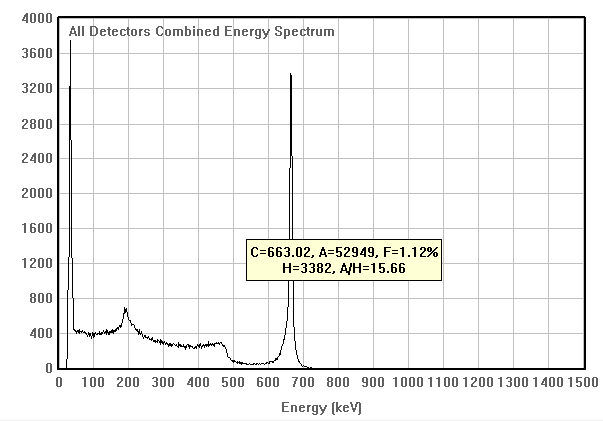}
         \caption{Post-calibration}
         \label{fig:monizPost}
     \end{subfigure}

        \caption{The unbiased detector Cs-137 spectrum (a) pre- and (b) post-calibration.}
        \label{fig:monizCalibration}

\end{figure}

On day 50 of the room temperature recovery period, the system underwent a full calibration by H3D, Inc. With the new calibration, the biased detector's resolution improved from $1.6\%$ to $1.23\%$ FWHM at $662 \ \mathrm{keV}$, shown in Fig.~\ref{fig:celineCalibration}, while Fig.~\ref{fig:monizCalibration} shows the unbiased detector's improvement from $1.4\%$ to $1.12\%$. The dramatic improvement is largely due to the detector's 3D position-sensitive capabilities which allow for a voxel-by-voxel energy correction.

\section{Activation Analysis}
\label{sec:activiation}

\begin{figure}[h!]
  \centering
  \includegraphics[trim={0cm 0cm 0cm 0cm}, clip, width=0.75\linewidth]{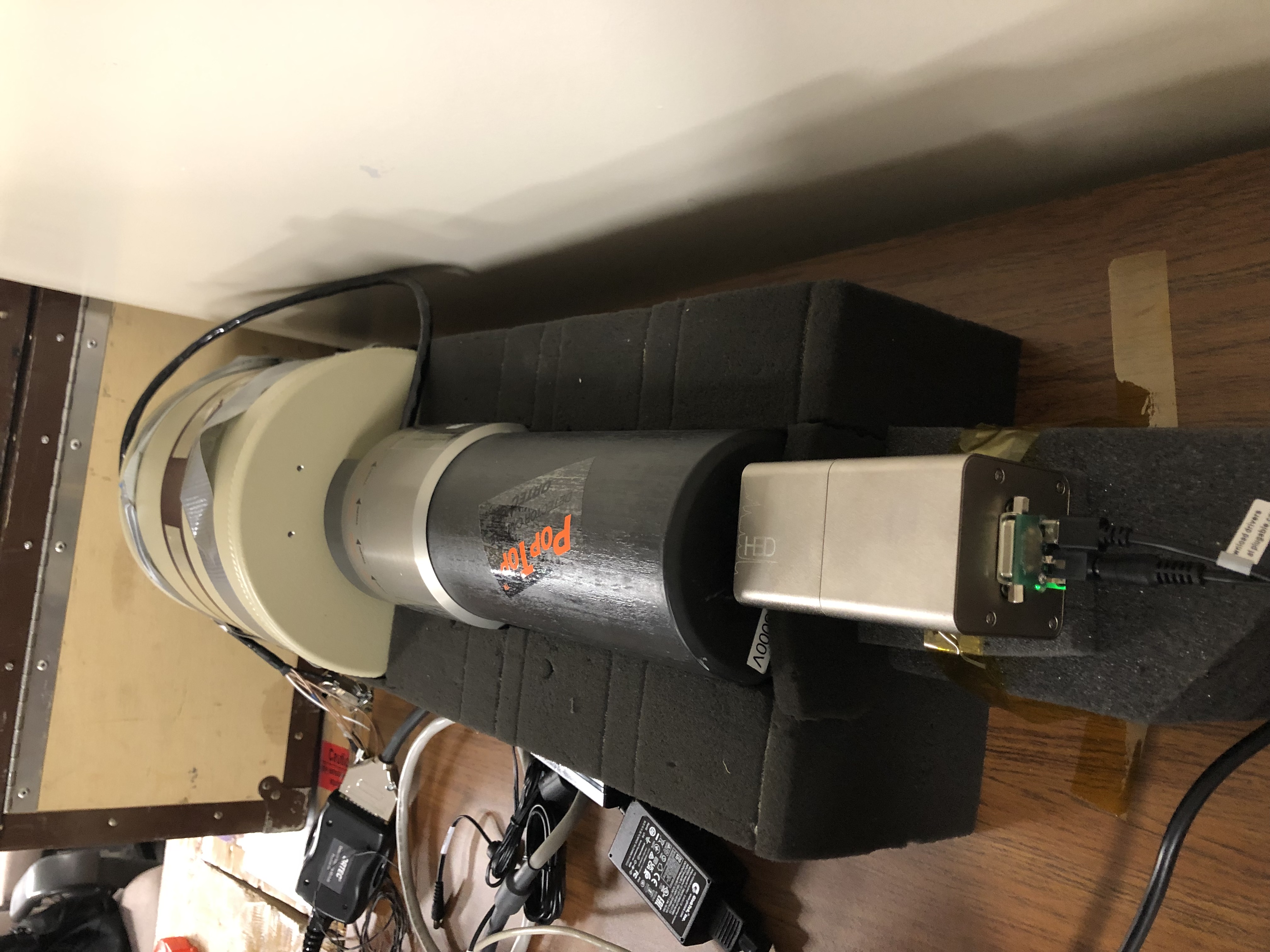}
  \caption{Activation study setup with the CZT detector centered on the front face of the co-axial HPGe detector.}
  \label{fig:crystalSetup}
\end{figure}

\begin{figure*}[h!]
  \centering
  \includegraphics[trim={0cm 0cm 0cm 0cm}, clip, width=0.85\textwidth]{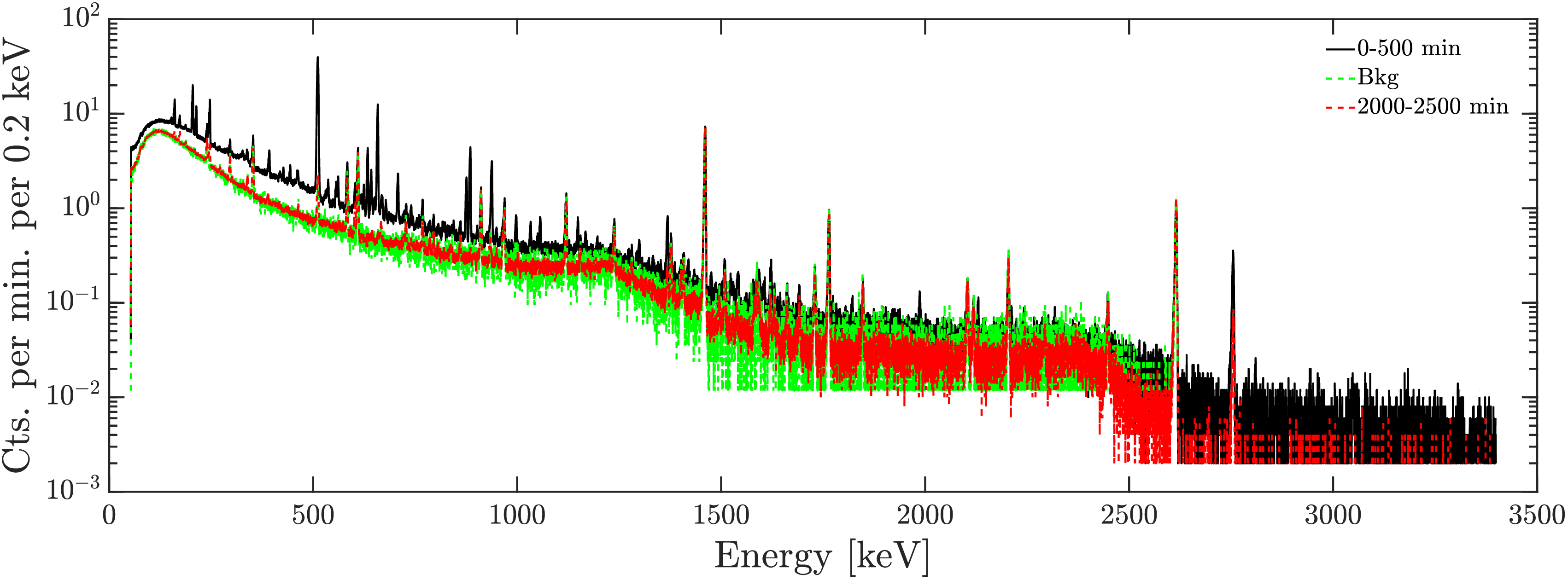}
  \caption{Full HPGe spectrum of the CZT detector taken after irradiation.}
  \label{fig:timeSpectra1d}
\end{figure*}

Immediately following the final irradiation, the CZT detectors were placed on the front face of a high-purity germanium (HPGe) detector to measure its activation products over two days. Fig.~\ref{fig:crystalSetup} shows the experimental setup. Fig.~\ref{fig:timeSpectra1d} shows the recorded spectra for the first 500 minutes (black solid line) and the last 500 minutes which was 2000 minutes into the measurements (dashed red line). The green dashed line represents a background measurement. The last 500 minutes show a significant reduction in internal activation with only a minor number of longer-lived products.

Table~\ref{tab:activation} lists the identified peak with the aid of InterSpec~\cite{interspec}, with similar observation by~\cite{activation1, BNLProton, cdProtonXSection}. The appendix of this manuscript contains additional HPGe and CZT spectral plots of the activation. The table omits typical background lines, except for 24Na lines, which is likely the result of the activated aluminum housing, as commonly observed in space missions~\cite{COMPTEL_Bkg}.

\begin{table*}[h!]
\caption{Identified radioisotopes in the activation spectrum taken by the HPGe detector. Along with the isotope, we report the energy, branching ratio, and half-life. The units for the half-life have the following short hand: m-minutes, h-hours, and d-days.}
\label{tab:activation}
\begin{minipage}[b]{0.45\linewidth}\centering
\begin{tabular}[t]{l l l l}
	\toprule
    Isotope & Energy [keV] & Branch Ratio [$\%$] & Half Life \\
    \hline
    \hline
Ag104 & 555.8 & 92.6 & 69.2m \\ \hline
        Cd-111m & 150.82 & 29.14 & 48.5m \\ 
         & 254.4 & 94.0 & 48.5m \\ \hline
        I-121 & 212.2 & 84.3 & 127.2m \\ \hline
        I-123 & 158.97 & 83.3 & 13.22h \\ \hline
        I-124 & 602.73 & 62.9 & 4.18d \\ 
         & 722.78 & 10.36 &  \\ 
         & 1690.96 & 11.15 & \\ \hline
        I-126 & 666.33 & 32.88 & 12.93d \\ 
         & 753.82 & 4.147 &  \\ \hline
        I-128 & 442.9 & 12.62 & 24.99m \\ \hline
        I-130 & 417.93 & 34.15 & 12.36h \\ 
         & 536.07 & 99.0 &  \\ 
         & 668.54 & 96.03 &  \\ 
         & 739.51 & 82.17 &  \\ \hline
        In-107 & 204.95 & 47.2 & 32.4m \\ \hline
        In-108 & 1032.8 & 35.35 & 58.0m \\ 
         & 1056.6 & 28.77 &  \\ 
         & 1299.3 & 15.07 &  \\ 
         & 1485.8 & 4.384 &  \\ 
         & 1606.3 & 8.494 &  \\ \hline
        In-108m & 632.9 & 76.4 & 39.6m \\ 
         & 875.4 & 2.445 &  \\ 
         & 1529.4 & 7.334 & \\ 
         & 1986.3 & 12.38 & \\ \hline
        In-109 & 203.5 & 73.5 & 4.17h \\ 
         & 288.4 & 1.617 &  \\ 
         & 347.5 & 2.131 &  \\ 
         & 426.2 & 4.116 &  \\ 
         & 613.6 & 2.278 &  \\ 
         & 619.0 & 1.764 &  \\ 
         & 623.5 & 5.513 & \\ 
         & 649.8 & 1.604 & \\ 
         & 949.1 & 1.313 & \\ 
         & 1049.7 & 1.169 & \\ 
         & 1149.1 & 4.337 & \\ 
         & 1196.5 & 1.617 & \\ 
         & 1272.9 & 0.5586 & \\ 
         & 1419.2 & 1.264 & \\ 
         & 1475.8 & 0.441 & \\ 
         \bottomrule
\end{tabular}
\end{minipage}
\hspace{0.75cm}
\begin{minipage}[b]{0.45\linewidth}
\centering
\begin{tabular}[t]{l l l l}
	\toprule
    Isotope & Energy [keV] & Branch Ratio [$\%$] & Half Life \\
    \hline
    \hline
In-110 & 461.8 & 4.718 & 4.9h \\ 
         & 560.32 & 1.867 & \\ 
         & 581.93 & 8.551 & \\ 
         & 584.21 & 6.487 & \\ 
         & 641.68 & 25.95 & \\ 
         & 657.75 & 98.29 & \\ 
         & 677.6 & 4.521 & \\ 
         & 707.4 & 29.49 & \\ 
         & 744.26 & 1.966 & \\ 
         & 759.87 & 3.145 & \\ 
         & 818.02 & 2.261 & \\ 
         & 844.67 & 3.244 & \\ 
         & 884.67 & 92.88 & \\ 
         & 901.53 & 1.966 & \\ 
         & 937.48 & 68.41 & \\ 
         & 997.16 & 10.52 & \\ \hline
        In-110m & 2129.4 & 2.158 & 69.1m \\ 
         & 2211.33 & 1.746 & \\ 
         & 2317.41 & 1.29 & \\ \hline
        In-111 & 171.28 & 90.65 & 67.31h \\ 
         & 245.35 & 94.08 & \\ \hline
        In-111m & 537.0 & 87.2 & 7.7m \\ \hline
        In-113m & 391.7 & 64.94 & 99.48m \\ \hline
        In-116m & 1293.4 & 84.8 & 54.29m \\ 
         & 1507.59 & 9.922 & \\ \hline
        Na-24 & 1368.63 & 100 & 14.9h \\ 
         & 2754.03 & 99.94 & \\ \hline
        Te-121 & 573.14 & 80.4 & 19.17d \\ 
        \bottomrule
\end{tabular}
\end{minipage}

\end{table*}

\section{Discussion}
\label{sec:discussion}

The shift in gain and degraded resolution observed during irradiation is likely a result of increased trapping sites due to bulk displacement damage which decreases the electron mobility lifetime product~\cite{Messenger1986TheEO}. As such, less charge induction occurs in damaged detectors as the average charge drifting through the weighting potential decreases. We observe the decrease of electron collection efficiency with the decrease in the 137Cs photopeak gain. Moreover, another result from Sec.~\ref{sec:damageResults} points to a possible bias dependence on radiation damage. This phenomenon is not new as other semiconductor-type devices observe this behavior~\cite{radEffectsOnSolidState, ccdRadDamage, biasDependanceSiDiodes, sipmBias}. A possible explanation is that the bias of the detector may promote the drift of interstitial vacancies whereas the unbiased case may only have thermal effects for random motion. This result is, however, not consistent with a previous result conducted by Franks et al. as they report less observed damage in the biased case. This is also contradictory to the behavior of other electronics.

The results also suggest that during orbit, to preserve the performance of the instrument, crystals should remain unbiased in the SAA. This is common in many space-based instruments.

We note that when we measured the 137Cs source in between irradiations, we placed the source on the cathode side. Since the Bragg peak occurred near the anode side, the majority of the damage occurred there. With a cathode side 137Cs placement, a lot of the events will interact on the cathode side. Should the source be located elsewhere, changing the distribution of gamma interactions, the spectrum will look different. A poly-chromatic proton beam could be used to spread out the Bragg peak to mitigate this effect (i.e. spread out Bragg peak)~\cite{portugalCZTRadDamage}.

The most dramatic recovery of the detector's performance occurred during the first two weeks. The performance recovery plateaued after 75 days with the unbiased detector's resolution returning to its original state while the biased did not. It took 78 hrs of $60 \ ^{\circ}\mathrm{C}$ annealing for the biased system's resolution to return to its original state. A full calibration of the system further improved the system's performance, owing to the 3D position sensing capability of the detector. Note that the applied dosages in this study is more acute than what would be experienced in spaceflight. Since spaceflight might be more chronic, the detectors will continue annealing.

\begin{figure}[]
  \centering
  \includegraphics[width=0.65\linewidth]{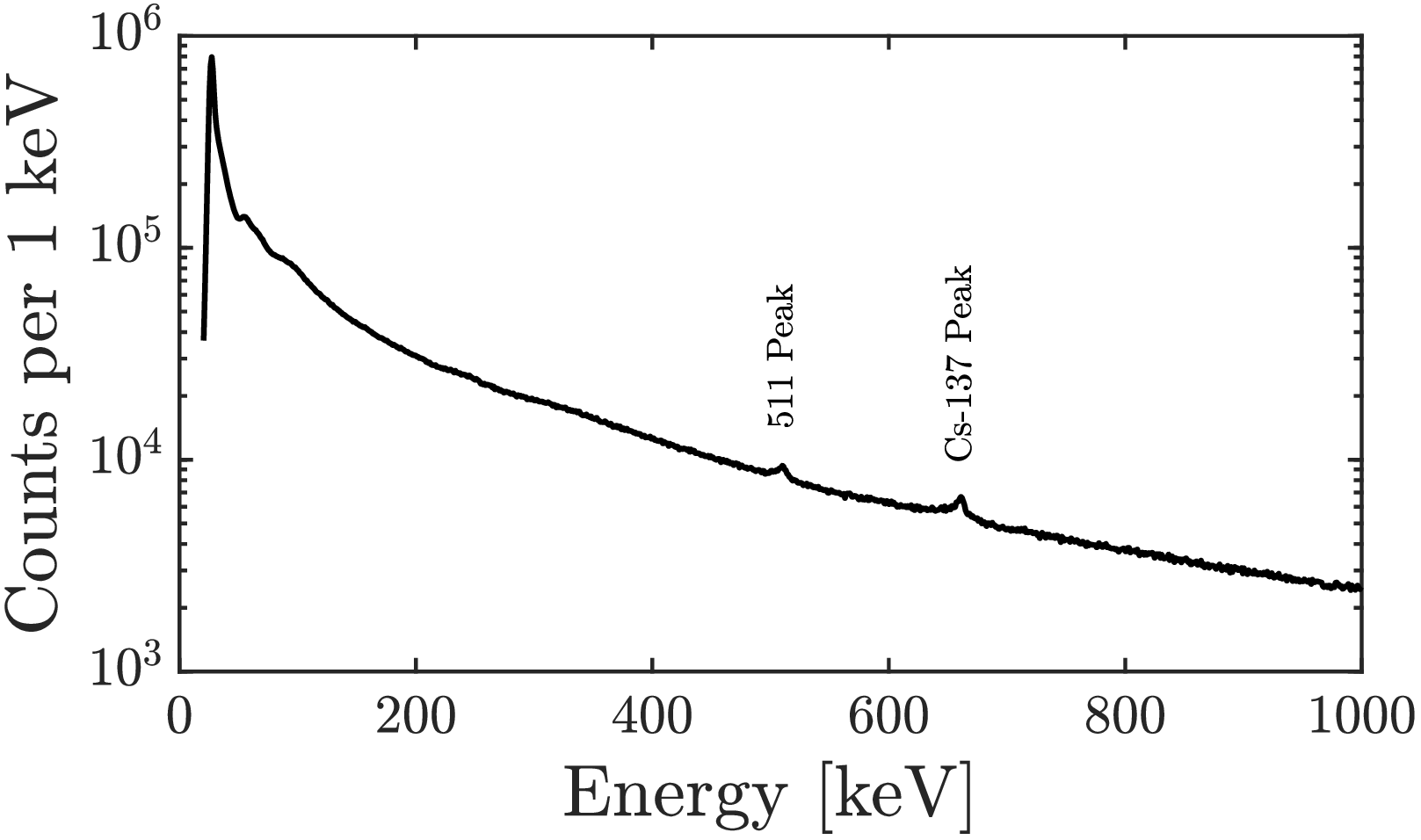}
  \caption{Spectra from when the detector was biased and acquiring during the first irradiation.}
  \label{fig:protonSpectra}
\end{figure}

With a significant deadtime, the powered M400 was able to remain somewhat unparalyzed during irradiation with a flux of $\sim10^5 \ \mathrm{p/cm^2/s}$. Fig.~\ref{fig:protonSpectra} plots the recorded spectra during the first irradiation of the biased detector. Both the 137Cs peak and a $511 \ \mathrm{keV}$ peak are visible along with a continuum. The results from those irradiations are still under investigation.

\section{Conclusion}
This work explores the radiation damage incurred on pixelated $2 \times 2 \times 1 \ \mathrm{cm}^3$ CZT detectors by $61 \ \mathrm{MeV}$ protons. The results suggest that CZT proton radiation hardness might correlate with the detector bias during irradiation with the biased detector degrading than  the unbiased detector. The biased detector began degrading, with an observable gain shift, with fluences as low as $8.7 \times 10^6 \ \mathrm{p/cm^2}$ while the resolution and centroid of the unbiased detector did not degrade a fluence of ${\sim} 8.8 \times 10^7 \ \mathrm{p/cm^2}$. The majority of proton induced degradation was recovered over a two week room temperature anneal. On day 50, a full system recalibration brought system resolutions down to $1.23 \%$ and $1.12 \%$ for the biased and unbiased detectors respectively. The recovery plateued after ${\sim}75$ days with the biased detector's 137Cs peak located at ${\sim}655.1 \ \mathrm{keV}$ with $1.6 \%$ FWHM while the unbiased detector was ${\sim}659.1 \ \mathrm{keV}$ with $0.9 \%$.  Further annealing at $60 \ ^{\circ} \mathrm{C}$ for a period of 78 hours allowed for the full resolution recovery of the system to its pre-irradiated state.

We are still investigating the response of the CZT detector when it was acquiring during irradiation. Finally, we are currently developing an instrument using $4 \times 4 \times 1.5 \ \mathrm{cm^3}$ pixelated CZT detectors with a slated launch date of early 2025 to the International Space Station. The project aims to space-qualify the large-volume crystals and associated electronics and study their operations in a space environment.

\section*{Acknowledgment}
We appreciate the efforts of Michael Backfish and the Crocker Nuclear Laboratory staff for ensuring a successful experiment. This work is supported by the Office of Naval Research 6.1. D. Shy is supported by the U.S. Naval Research Laboratory's Jerome and Isabella Karle Fellowship.

The H3D, Inc. portions of this work were funded under the Department of Energy Small Business Innovative Research Program award number DE-SC0021765.

\appendix

\section{CZT Measurement of Internal Activation}

\begin{figure*}[h!]
  \centering
  \includegraphics[width=0.6\textwidth]{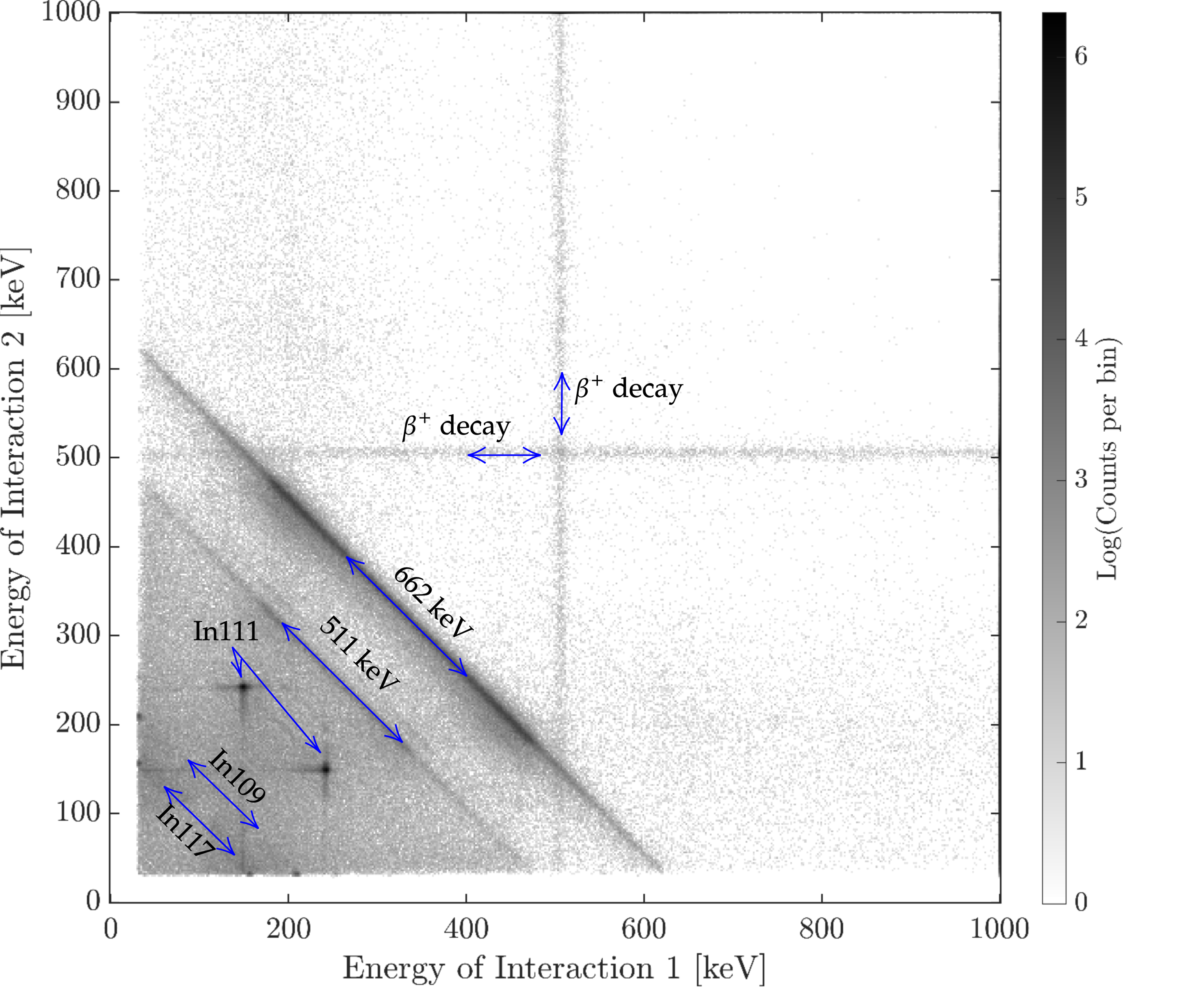}
  \caption{Bivariate plot of two-interaction events of a 137Cs measurement taken right after the full proton irradiation with the `biased' detector.}
  \label{fig:bivariateActivation}
\end{figure*}

The CZT system is also able to intrinsically measure internal activation. Fig.~\ref{fig:bivariateActivation} plots the response of two-interaction events for a 137Cs measurement taken right after irradiation. Since CZT does not have the timing resolution to distinguish between interactions, the temporal order of interactions cannot be resolved. The labeling of interactions 1 and 2 are therefore arbitrary and interchangeable. In the plot, diagonals such as those indicated $511$ and $662 \ \mathrm{keV}$, represent gamma rays that scatter and are subsequently absorbed in CZT. A portion of the horizontal and vertical events at $511 \ \mathrm{keV}$ as one of the interactions may be represented by a $\beta^+$ decay. There, the continuum represents the positron's kinetic energy, which will then annihilate to produce two $511 \ \mathrm{keV}$ photons. In this plot, one of the annihilation photons escaped. A coincident background event and a $511 \ \mathrm{keV}$ measurement could also contribute to the distribution. The two islands of events labeled as In111 originate from measuring the $171.3$ and $245.4 \ \mathrm{keV}$ photons simultaneously.

\begin{figure*}[h!]
  \centering
  \includegraphics[width=0.5\textwidth]{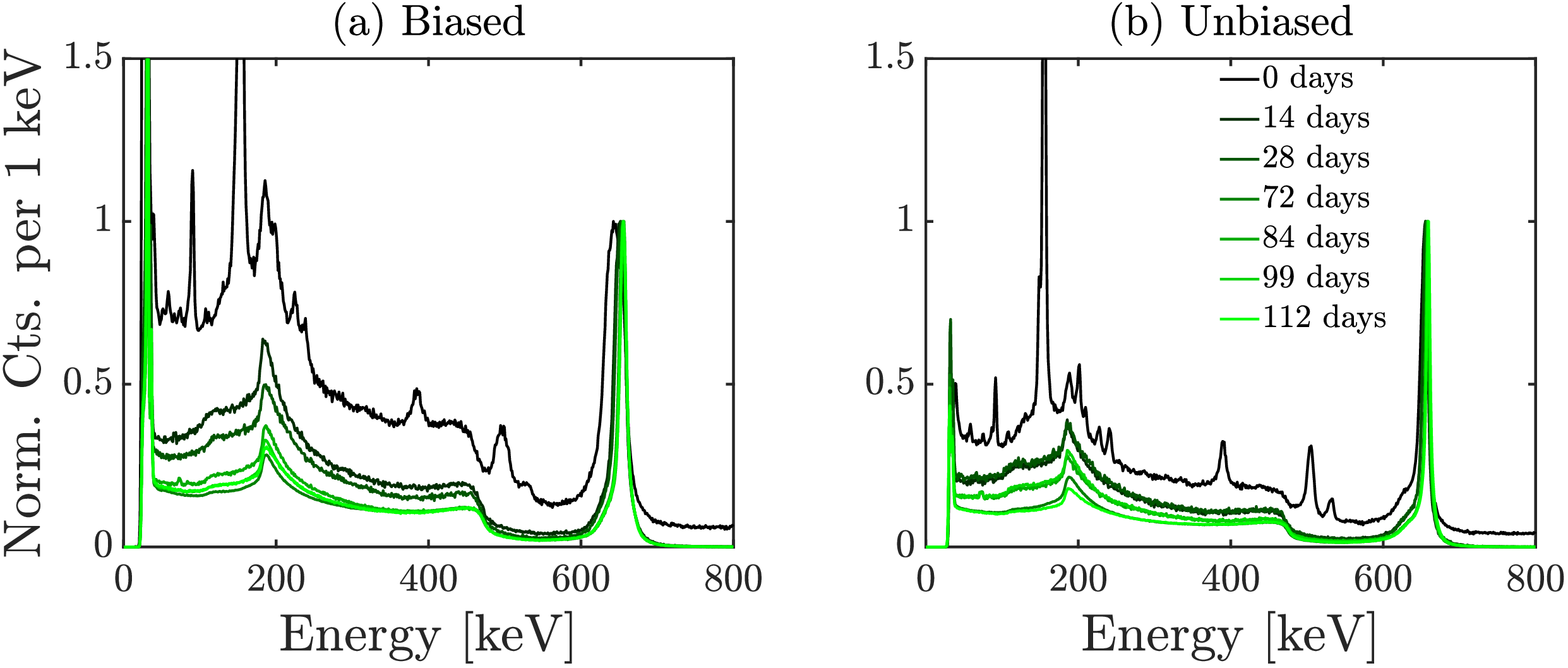}
  \caption{Change of spectrum over time following irradiation. Note that the 137Cs source location and environmental factors were not controlled between different measurements. This plot is the same as Fig.~\ref{fig:recoverySpectra}, just with a larger energy range to display the internal activation.}
  \label{fig:fullAnnelingCurve}
\end{figure*}

Finally, we plot the full gamma-ray spectrum of the 137Cs source over time in Fig.~\ref{fig:fullAnnelingCurve} as taken by the CZT detectors during the recovery study. After two weeks, the spectra show the majority of internal activation has decayed.

\section{Additional HPGe Plots}

\begin{figure*}[h!]
  \centering
  \includegraphics[trim={0cm 0cm 0cm 0cm}, clip, width=0.9\textwidth]{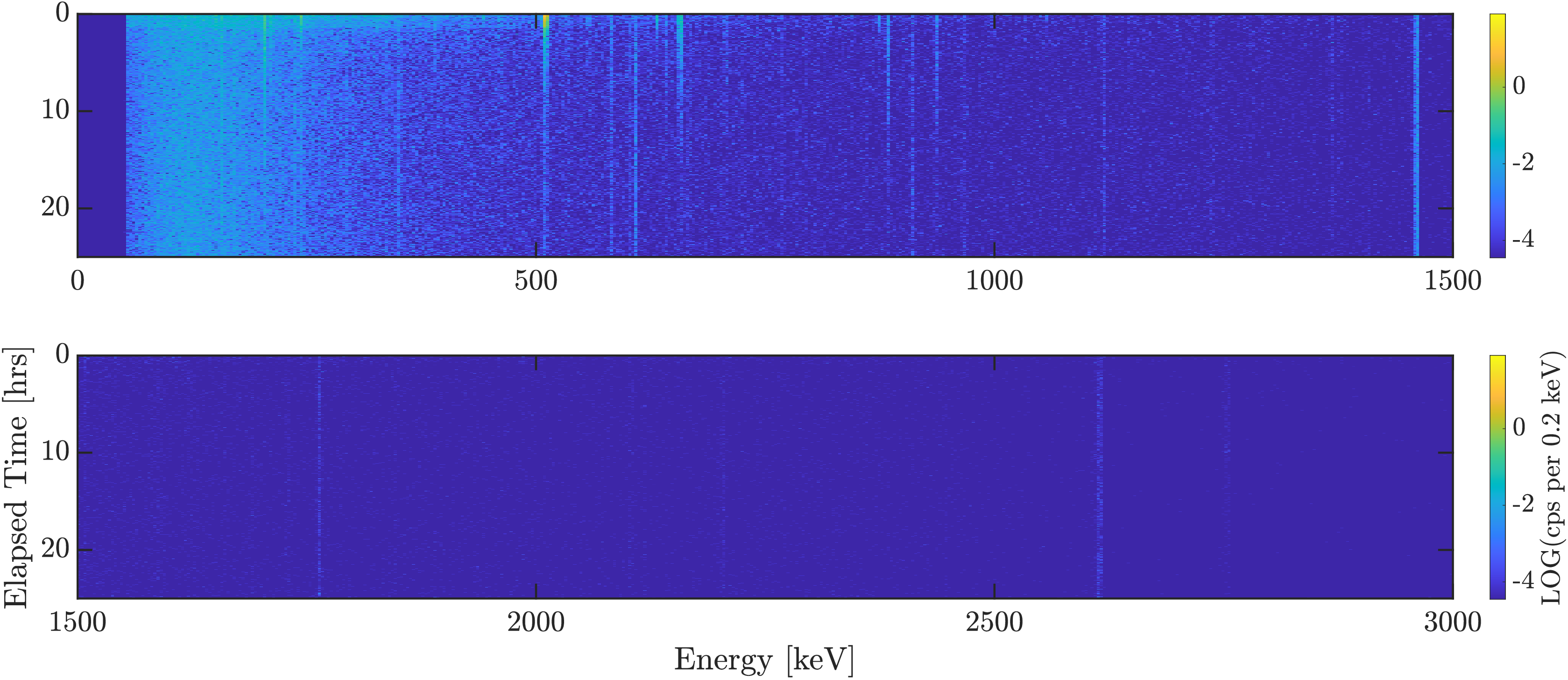}
  \caption{Spectrogram of the gamma-ray spectrum over time}
  \label{fig:timeSpectra}
\end{figure*}

This appendix offers additional plots taken by the HPGe to study the activation productions. Fig.~\ref{fig:timeSpectra} presents a waterfall plot showing the time evolution of the gamma-ray spectrum. The energy range is divided into two to highlight the different regions (top and bottom).

\begin{figure*}[h!]
  \centering
  \includegraphics[width=0.8\textwidth]{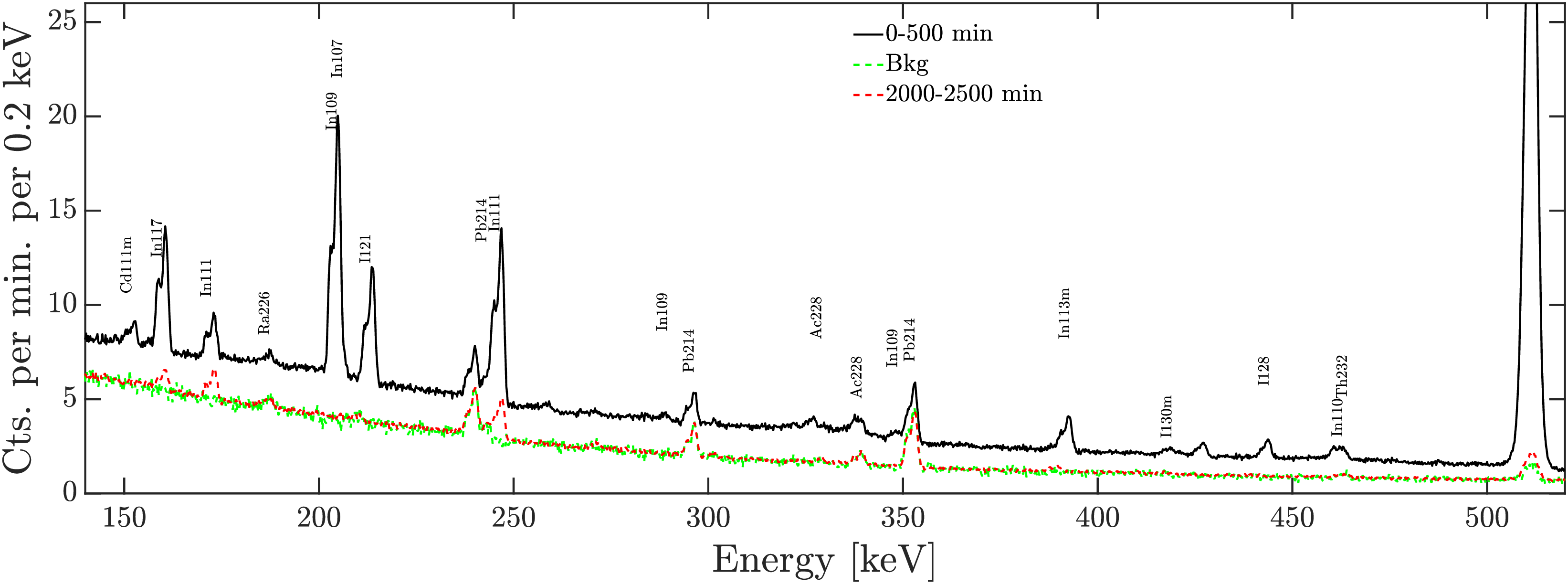}
  \caption{HPGe gamma-ray spectrum of the activated CZT ranging from $140$ to $520 \ \mathrm{keV}$}
  \label{fig:region0}
\end{figure*}

\begin{figure*}[h!]
  \centering
  \includegraphics[width=0.85\textwidth]{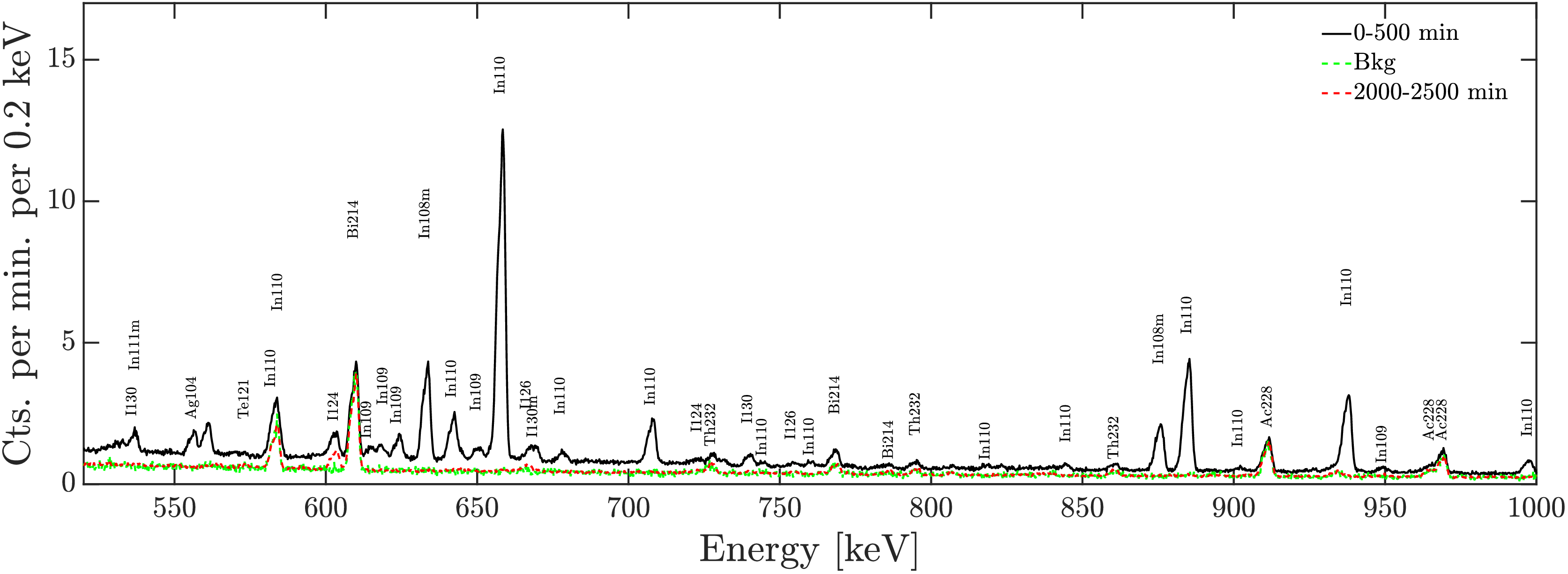}
  \caption{HPGe gamma-ray spectrum of the activated CZT ranging from $520$ to $1000 \ \mathrm{keV}$}
  \label{fig:region1}
\end{figure*}

\begin{figure*}[h!]
  \centering
  \includegraphics[width=0.85\textwidth]{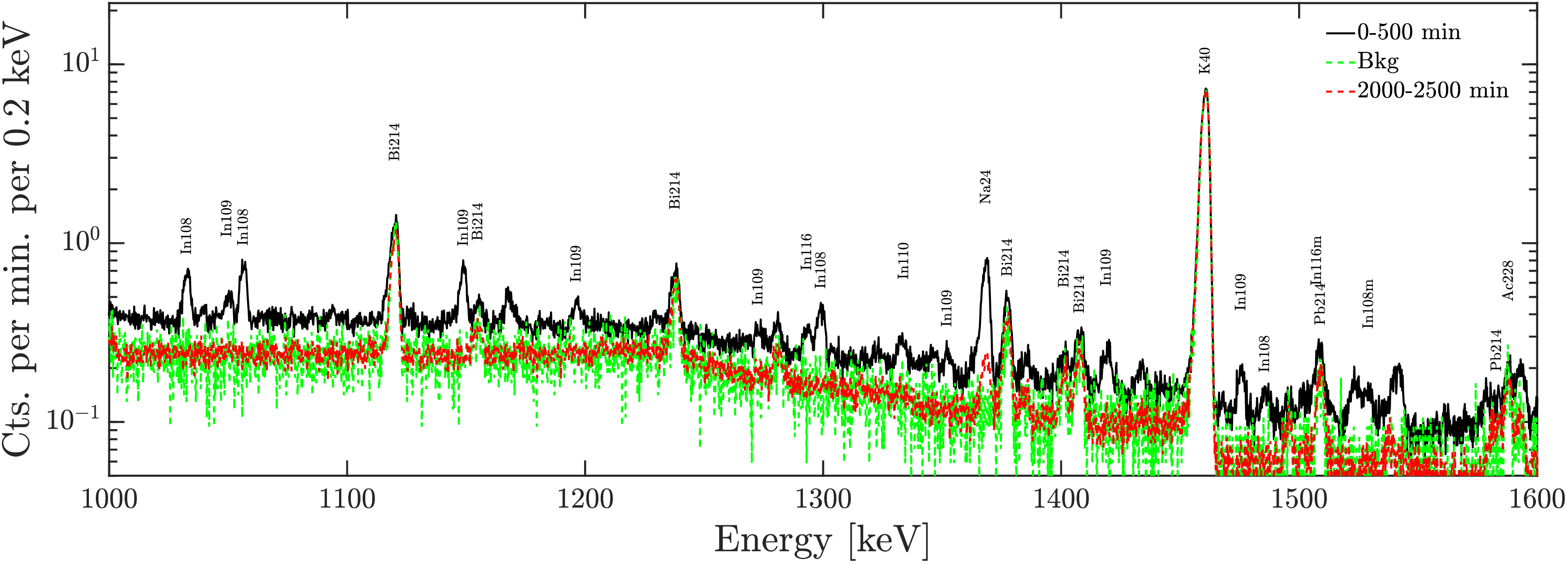}
  \caption{HPGe gamma-ray spectrum of the activated CZT ranging from $1000$ to $1600 \ \mathrm{keV}$}
  \label{fig:region2}
\end{figure*}

\begin{figure*}[h!]
  \centering
  \includegraphics[width=0.85\textwidth]{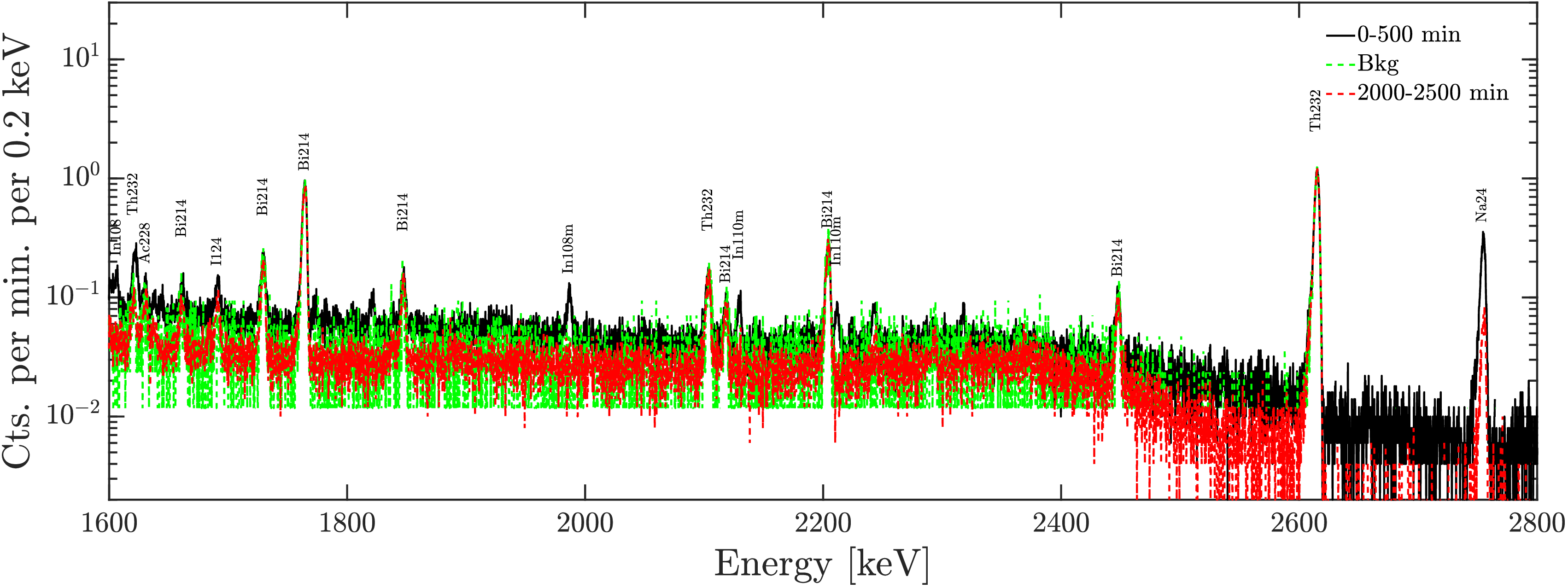}
  \caption{HPGe gamma-ray spectrum of the activated CZT ranging from $1600$ to $2800 \ \mathrm{keV}$}
  \label{fig:region3}
\end{figure*}

Figs.~\ref{fig:region0},~\ref{fig:region1},~\ref{fig:region2},~\ref{fig:region3} present the HPGe spectra with different energy ranges comparing the first 500 minutes, the gamma-ray background, and the last 500 minutes of the measurements which were 2000 minutes into the measurement.

\nocite{*}
\bibliographystyle{elsarticle-num}

\bibliography{IEEEbib}

\end{document}